\documentclass[twocolumn,showpacs,preprintnumbers,amsmath,amssymb,aps]{revtex4}
\usepackage{graphicx}
\usepackage{dcolumn}
\usepackage{bm}

\begin{document}

\title{Precise measurement of hyperfine structure in the ${2P}_{1/2}$ state
of $^{7}$Li using saturated-absorption spectroscopy}
 \author{Alok K. Singh, Lal Muanzuala, and Vasant Natarajan}
 \email{vasant@physics.iisc.ernet.in}
 \homepage{www.physics.iisc.ernet.in/~vasant}
 \affiliation{Department of Physics, Indian Institute of
 Science, Bangalore 560\,012, INDIA}

\begin{abstract}
We report a precise measurement of the hyperfine interval
in the ${2P}_{1/2}$ state of $^{7}$Li. The transition from
the ground state ($D_1$ line) is accessed using a diode
laser and the technique of saturated-absorption
spectroscopy in hot Li vapor. The interval is measured by
locking an acousto-optic modulator to the frequency
difference between the two hyperfine peaks. The measured
interval of 92.040(6)~MHz is consistent with an earlier
measurement reported by us using an atomic-beam
spectrometer [Das and Natarajan, J.\ Phys.\ B {\bf 41},
035001 (2008)]. The interval yields the magnetic dipole
constant in the $P_{1/2}$ state as $A=46.047(3)$, which is
discrepant from theoretical calculations by $>80$~kHz.
\end{abstract}



\maketitle

\section{Introduction}
Measurement of hyperfine structure in the low-lying states
of Li is motivated by the fact that its simple
three-electron structure makes it amenable to accurate
calculations. We have recently reported the most-precise
measurements to date of hyperfine intervals (with accuracy
of 6~kHz) in the ${2P}_{1/2}$ state of $^{6,7}$Li
\cite{DAN08}. Around the same time, there have been
high-accuracy calculations of the hyperfine constants in
this state using both the configuration-interaction method
\cite{YER08} and the coupled-cluster method \cite{DPB08}.
Furthermore, Beloy and Derevianko \cite{BED08} have pointed
out the importance of taking second-order effects into
account when calculating the hyperfine constants from the
measured intervals. The largest corrections are for Li, and
in the case of $^{7}$Li, the corrections are much larger
than the experimental error.

The measured interval in $^{6}$Li reported by us in the
earlier work was 26.091(6)~MHz. From this, we obtained an
uncorrected constant of $A=17.394(4)$~MHz \cite{DAN08}.
However, taking the $+4.01$~kHz second-order correction
given in Ref.\ \cite{BED08}, the measured interval yields a
constant of 17.398(4)~MHz, which is in good agreement with
the calculated value of $17.4058(8)$~MHz reported in Ref.\
\cite{YER08}. However, the situation in $^{7}$Li is quite
unsatisfactory. Our measured interval of 92.047(6)~MHz
yields an uncorrected constant of $A=46.024(3)$~MHz. The
second-order correction from Ref.\ \cite{BED08} is $+27.0$
kHz, almost an order-of-magnitude larger than the
experimental error. More importantly, there are two
independent calculations of the hyperfine constant, one
reporting a value of $45.966(1)$~MHz \cite{YER08} and the
other a value of $45.958$~MHz \cite{DPB08}. Thus, the
uncorrected constant from our measurement differs from both
calculations by $\sim 60$~kHz ($20 \, \sigma$), while the
correction increases the difference still further. Indeed,
the agreement in $^6$Li and the discrepancy in $^7$Li is
surprising because one would expect larger errors in $^6$Li
due to its $13 \times$ smaller natural abundance, resulting
in correspondingly lower signal-to-noise ratio.

Though theoretically Li is the simplest alkali-metal atom
to deal with, experimentally it is the most challenging.
This is because of its high reactivity with all kinds of
glasses, which precludes the use of vapor cells. Vapor
cells exist for all the other alkali-metal atoms, and the
standard technique of saturated-absorption spectroscopy
(SAS) \cite{DEM82} in a vapor cell gives spectra with
linewidths close to the natural linewidth. This problem can
be partly addressed by using a buffer gas in the cell, but
that causes collisional broadening of the line and large
{\it shifts} of the line center \cite{CAR07}, making it
useless for precision measurements. As a result, most
high-precision spectroscopy experiments in Li have been
done using laser-induced fluorescence from a collimated
atomic beam. In such experiments, it is crucial that the
laser beam be perpendicular to the atomic beam. Any
misalignment angle would cause a systematic Doppler shift
of line center. Since our earlier experiments on Li
\cite{DAN08} were done with an atomic beam, and because of
the discrepancy with theoretical calculations of the
hyperfine constant in $^7$Li, we decided to repeat these
measurements using the SAS technique in a new
non-atomic-beam spectrometer. Most important for precision
measurements is that the SAS technique does not cause a
systematic shift of line center even if there is a small
misalignment angle between the pump and probe beams.

In this work, we present results of measurements of the
hyperfine interval in the $2P_{1/2}$ state of $^7$Li using
{\it saturated-absorption spectroscopy} in this new
spectrometer. To the best of our knowledge, this is the
first time such high-resolution SAS spectra have been
obtained in Li. The interval is measured with our
well-developed technique of locking an acousto-optic
modulator (AOM) to the frequency difference between two
hyperfine peaks \cite{DAN08,RKN03}. The results from the
current set of measurements are consistent with the
previous set, giving confidence that Doppler-shift errors
were under control in the previous atomic-beam
spectrometer.

\section{Experimental details}
The new Li spectrometer consisted of a 25-mm-diameter
$\times$ 100-mm-long cylindrical pyrex cell. The cell was
connected to a resistively-heated Li source and a
turbo-molecular pump to maintain pressures below $10^{-7}$
torr. During the experiment, the cell was continuously
loaded with Li vapor by heating the source. If the source
was turned off, the pump would rapidly evacuate the
remaining atoms. The ${2S}_{1/2} \rightarrow {2P}_{1/2}$
transition ($D_1$ line) in Li is at 670 nm. This was
accessed with a home-built diode laser system \cite{BRW01},
which was frequency stabilized using grating feedback to
give an rms linewidth of 1~MHz. The two beams were elliptic
with size of 4 mm $\times$ 1.5 mm. The probe beam had a
power of 12 $\mu$W and the pump power was varied from 50 to
120 $\mu$W. Thus the highest pump intensity was less than
the saturation intensity of 2.5 mW/cm$^2$.

\begin{figure}
\centering{\resizebox{0.9\columnwidth}{!}{\includegraphics{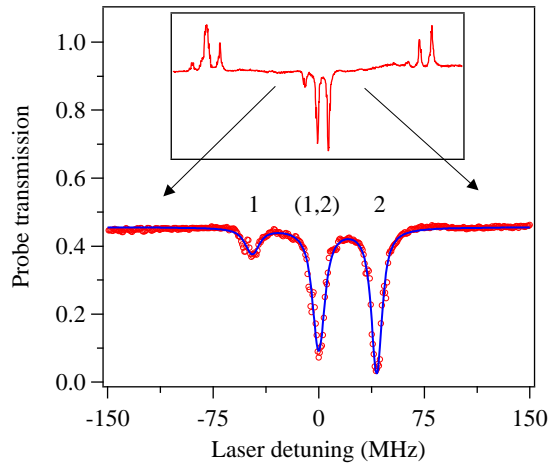}}}
\caption{(Color online) Saturated-absorption spectrum in
$^7$Li. The inset is the complete $F \rightarrow F'$
spectrum for the $D_1$ line showing respectively
transitions starting from the $F=1$ ground level, due to
the $F=(1,2)$ ground crossover resonance, and transitions
from the $F=2$ ground level. Each set has three peaks
corresponding to $F'=1$, $F'=(1,2)$ crossover resonance,
and $F'=2$. The main spectrum is a close-up of the ground
crossover resonance with peaks labeled with the value of
$F'$. The solid curve is a multipeak fit to a Voigt
profile.}
 \label{spec1}
\end{figure}

A typical spectrum of the $D_1$ line in $^7$Li is shown in
Fig.\ \ref{spec1}. The spectrum is Doppler corrected by
subtracting the signal from a second probe beam without a
counter-propagating pump beam. The entire spectrum, as
shown in the inset of the figure, has three sets of peaks.
The first and third set are easy to understand; they are
for transitions starting from the $F=1$ ground level and
the $F=2$ ground level respectively. It is well known that
the use of counter-propagating pump-probe beams in SAS
causes spurious crossover resonances, exactly in between
two real peaks. The real peaks are caused when the pump
saturates the transition for zero-velocity atoms, so that
probe absorption is reduced. The crossover resonances are
caused when a non-zero-velocity group is resonant with one
transition for the probe beam and the other transition for
the pump beam. Thus two velocity groups contribute to each
crossover resonance, and these resonances are generally
more prominent than the real peaks. As the figure inset
shows, each set of transitions has three peaks
corresponding to $F'=1$, $F'=(1,2)$ (the prominent
crossover resonance), and $F'=2$.

The middle set of peaks is a further artifact of SAS, and
is called a ground crossover resonance labeled as $F=(1,2)
\rightarrow F'$. It appears because the ground hyperfine
interval in $^7$Li is 803~MHz, which corresponds to a
Doppler shift within the velocity profile of the Li vapor.
Thus, for atoms moving with a velocity $v=+269$~m/s and
when the laser frequency is exactly between transitions
starting from the $F=1$ and $F=2$ levels, the pump induces
transitions from the $F=1$ level and optically pumps atoms
into the $F=2$ level. For the same velocity group, since
the counter-propagating probe beam (with opposite Doppler
shift) is exactly resonant with transitions starting from
the $F=2$ level, the probe beam shows {\it enhanced
absorption}. The roles of the $F=1$ and $F=2$ levels are
interchanged for atoms moving with $v=-269$~m/s. The ground
crossover set is therefore inverted compared to the other
two sets (see figure) and, as expected, more prominent. We
have therefore used this peak for the measurements. A
multipeak fit to this set of peaks with a Voigt profile
fits the spectrum quite well and shows that there is no
significant asymmetry of the line shape. The
full-width-at-half-maximum of each peak is about 10~MHz.
The Voigt profile and linewidth can be understood as
arising from a combination of a Lorentzian contribution due
to the natural linewidth of 6~MHz and a Gaussian
contribution from a small angle between the
counter-propagating beams.

The measurement of the hyperfine interval proceeds as
follows. The laser is first locked to the $F=(1,2)
\rightarrow F'=2$ transition using SAS in the vapor cell.
For locking, the diode current is modulated at $f=20$~kHz
with a depth of modulation of about 2~MHz, and the signal
demodulated at $3f$ is fed back to the piezoelectric
transducer controlling the angle of the diode-laser
grating. Such third-harmonic locking makes the locking
insensitive to the underlying Doppler profile in the probe
spectrum \cite{WAL72}. The locked beam is sent through
another part of the cell for a second SAS. Here, we use a
counter-propagating pump beam whose frequency is shifted
and scanned using an AOM, so that it is nearly resonant
with the $F=(1,2) \rightarrow F'=1$ transition. Scanning
only the pump beam is a technique that we have developed to
make the spectrum appear on a Doppler-free background
\cite{BAN03}. In other words, the locked probe beam
addresses only those atoms which are moving at $v=\pm
269$~m/s, so that it drives transitions to the $F'=2$
level. This absorption remains flat until the pump beam
also comes into resonance with the same $v=\pm 269$~m/s
atoms, which drives transitions to the $F'=1$ level and
increases probe absorption due to optical pumping (as
explained earlier). The frequency shift that brings the
pump beam into resonance is exactly the hyperfine interval
between the $F'=2$ and $F'=1$ levels. Thus, by feeding back
the demodulated signal at $f$ from the second SAS to the
AOM driver, we can lock its frequency to the interval,
which in turn can be read using a frequency counter.

\section{Error Analysis}
The different sources of error in the technique have been
discussed extensively in our earlier publications
\cite{DAN08,RKN03}, and are reviewed here for completeness.

\subsection{Statistical errors}
The primary sources of statistical error are the
fluctuations in the lock point of the laser and the AOM. To
minimize this, we integrate for 10 s for each reading of
the frequency counter driving the AOM. We then take an
average of about 10 independent measurements. The timebase
in the frequency counter has a stability of better than
$10^{-6}$, which translates to a negligible error of $<0.1$
kHz in the frequency measurement.

\subsection{Systematic errors}
Systematic errors can occur if there are systematic shifts
in the lock points of the laser and the AOM. This can arise
due to one of the following reasons.
\begin{enumerate}
\item[(i)] {\it Radiation-pressure effects.} Radiation
    pressure causes velocity redistribution of the
    atoms in the vapor cell. In the SAS technique, the
    opposite Doppler shifts for the counter-propagating
    beams can result in asymmetry of the observed
    lineshape. We minimize these effects by using beam
    intensities that are smaller than the saturation
    intensity, and the spectrum in Fig.\ \ref{spec1}
    shows that the observed line shape is symmetric.

\item[(ii)] {\it Effect of stray magnetic fields.} The
    primary effect of a magnetic field is to split the
    Zeeman sublevels and broaden the line without
    affecting the line center. However, line shifts can
    occur if there is asymmetric optical pumping into
    Zeeman sublevels. For a transition $|F,m_F \rangle
    \rightarrow |F',m_{F'} \rangle$, the systematic
    shift of the line center is $\mu_B(g_{F'}m_{F'} -
    g_Fm_F)B$, where $\mu_B=1.4$~MHz/G is the Bohr
    magneton, $g$'s denote the Land\'e $g$ factors of
    the two levels, and $B$ is the magnetic field. The
    selection rule for dipole transitions is $\Delta m
    = 0,\pm 1$, depending on the direction of the
    magnetic field and the polarization of the light.
    Thus, if the beams are linearly polarized, there
    will be no asymmetric driving and the line center
    will not be shifted. We therefore minimize this
    error by using polarizing beam-splitter cubes to
    ensure that the beams have near-perfect linear
    polarization.

\item[(iii)] {\it Phase shifts in the feedback loop.}
    We check for this error by replacing the AOM with
    two identical AOMs, and adjusting them so that they
    produce opposite frequency offsets. With the laser
    locked to a given hyperfine transition, the first
    AOM then produces a fixed frequency offset which is
    compensated by the second AOM. Thus the same
    hyperfine transition is used for locking in both
    spectrometers. Under these conditions, the second
    AOM should lock to the fixed frequency of the first
    AOM, with any error arising solely due to
    phase-shift errors. We find that the second AOM
    tracks the frequency of the first AOM to within
    1~kHz.

\item[(iv)] {\it Shifts due to collisions.} To first
    order, collisional shifts are the same for
    different hyperfine levels, and hence do not affect
    the interval. Small differential shifts of the
    interval have been studied carefully in the ground
    state of Cs, due to its importance in atomic
    clocks. However, the size of the shift is in the
    mHz range. Collisional shifts are also important in
    buffer-gas filled cells.

\item[(v)] {\it Peak pulling.} Though the two peaks
    that we lock to are 94~MHz apart (compared to the
    linewidth of 10~MHz), we have to consider that the
    locations of the $3f$ lock points will be pulled by
    the central peak in Fig.\ \ref{spec1}.

\end{enumerate}

The sizes of the various sources of error are listed in
Table \ref{t1}. Since the laser and AOM are locked, the
linearity of the laser-scan axis in Fig.\ \ref{spec1} is
not important. As mentioned in point (i) above, the peak
center can be shifted due to radiation-pressure effects.
Similarly, from point (ii), the center can be shifted if
there is {\it asymmetric} pumping into the Zeeman sublevels
in the presence of a residual magnetic field. Asymmetric
optical pumping can occur if the beam polarizations are not
perfectly linear, for example due to imperfections in the
cubes or birefringence at the cell windows. From an
experimental point of view, both these effects will change
with laser power. Therefore, we can check for our estimate
of these errors by repeating the measurements at different
values of power and extrapolate to zero power.

\begin{table}
\caption{Error budget.}
\begin{ruledtabular}
\begin{tabular}{lc}
\hspace*{3mm} Source of error & Size (kHz) \\
\hline
1. Optical pumping into Zeeman sublevels & 5 \\
2. Feedback loop phase shift & 2 \\
3. Collisional shifts & 2 \\
4. Peak pulling & 2
\end{tabular}
\end{ruledtabular}
 \label{t1}
\end{table}

\section{Results}
\begin{figure}
\centering{\resizebox{0.9\columnwidth}{!}{\includegraphics{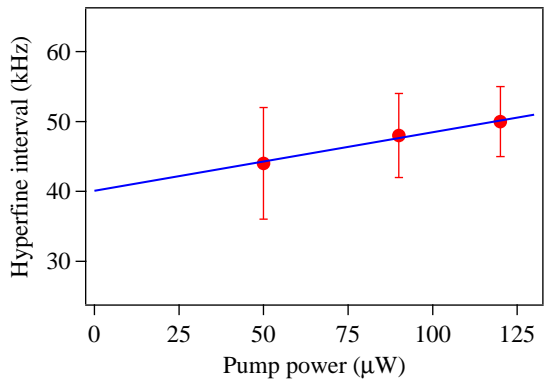}}}
\caption{(Color online) Pump power dependence of the
measured hyperfine interval. For simplicity, a fixed offset
of 92~MHz has been removed. The error bars represent the
standard deviation in each set. The solid line is a
weighted linear fit extrapolated to zero power.}
 \label{pow1}
\end{figure}

The results of measurements at different pump powers is
shown in Fig.\ \ref{pow1}. The vertical error bars give the
standard deviation in each set. Although the power is
increased by a factor of 2.5, the interval only changes by
2~kHz, which is less than the error bars. The straight-line
fit yields a zero-power $y$ intercept of 40.1(1)~kHz. By
adding in quadrature the different sources of error in
Table \ref{t1}, we estimate the total error in the average
value to be 6~kHz. Thus our current value for the interval
is
\begin{tabbing}
\hspace{0.75cm} $^{7}$Li, $2P_{1/2}$: \= \;$\Delta \nu_{2 -
1} = 92.040(6)$~MHz.
\end{tabbing}
This value can be compared to our previous measurement done
with the atomic-beam spectrometer, 92.047(6)~MHz. Within
their error bars, both results are consistent with each
other, though they have been done with completely different
spectroscopy techniques. From the interval, we obtain the
experimental value of the magnetic dipole constant $A$ as
46.047(3)~MHz, taking into account the $+27.0$~kHz
second-order correction from Ref.\ \cite{BED08}.

\section{Conclusion}
In conclusion, we have repeated the measurement of the
hyperfine interval in the $2P_{1/2}$ state of Li using the
SAS technique. As seen in Fig.\ \ref{spec1}, this technique
produces spurious crossover resonances that are more
prominent than the real peaks. Therefore, it can only be
used in cases where the interval between hyperfine levels
is at least 3 to 4 times the linewidth. Since the observed
linewidth in Li is 10~MHz, the SAS technique does not give
resolved peaks in the $D_1$ line of $^6$Li (where the
interval in the $2P_{1/2}$ state is only 26~MHz). However,
we have seen that the discrepancy with theory is only in
the case of $^7$Li. Indeed, the large discrepancy in $^7$Li
and the agreement in $^6$Li is surprising because the
signal-to-noise ratio in $^6$Li is 13 times worse due to
its smaller natural abundance (7.3\% vs.\ 92.7\%), and
because velocity-dependent errors will be larger for the
lighter isotope.

\begin{figure}
\centering{\resizebox{0.9\columnwidth}{!}{\includegraphics{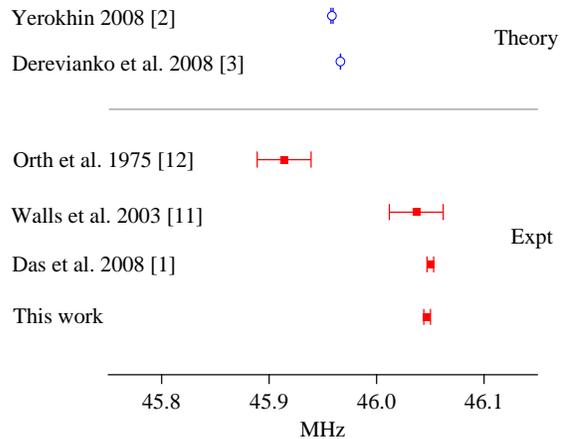}}}
\caption{(Color online) Comparison of experimental and
theoretical values of the hyperfine constant $A$ in the
$P_{1/2}$ state of $^7$Li. The three most recent
experimental values are corrected for second-order effects
from Ref.\ \cite{BED08}.}
 \label{compare}
\end{figure}

The current situation for the magnetic dipole constant $A$
in $^7$Li is summarized in Fig.\ \ref{compare}. The present
study yields a value of 46.047(3)~MHz, which is consistent
with a previous value from our laboratory \cite{DAN08} and
with another experiment \cite{WAC03}. Both these previous
experiments measured the hyperfine interval directly, and
the hyperfine constant was extracted from the interval. The
reported values have now been corrected for the
second-order effects given in Ref.\ \cite{BED08}. The
figure clearly shows that all the three recent experimental
values are discrepant with the two theoretical values.
There is an older measurement by Orth et al.\ \cite{OAO75}
from 1975, where a magnetic field was applied to resolve
the transitions and a complex fitting routine was used to
extract the hyperfine constants from the observed
optical-double-resonance and level-crossing signals. That
value is discrepant from the other experiments by more than
$3.5 \, \sigma$, and on the other side of theory at the $2
\, \sigma$ level.

Recently, we have also completed measurements of the
fine-structure splitting and isotope shifts in Li using the
atomic-beam spectrometer \cite{DAN07}. Here again, there
are fairly significant discrepancies with theory for the
$D_2$ isotope shift and the splitting isotope shift (SIS)
\cite{YND08}. We plan to repeat these measurements with the
new spectrometer. Though the peaks in the $D_2$ line are
not resolved (because the intervals are smaller than the
natural linewidth), we plan to use a magnetic field to
separate them.

\section*{Acknowledgments}
This work was supported by the Department of Science and
Technology, India. A.K.S. acknowledges financial support
from the Council of Scientific and Industrial Research,
India.


\end{document}